\documentclass[print]{raa}            

\usepackage{graphicx,times}             
\usepackage{natbib}
\usepackage{amssymb,amsmath}
\bibpunct{(}{)}{;}{a}{}{,}

\usepackage{CJKutf8}
\usepackage{comment}
\usepackage{color}
\usepackage{footmisc}

\newcommand{\beq}{\begin{eqnarray}}
\newcommand{\eeq}{\end{eqnarray}}

\newcommand{\fa}{f_{A}}
\newcommand{\fac}{f_{A,2}}
\newcommand{\Gc}{\Gamma_2}
\newcommand{\ffrb}{f_{\rm FRB}}

\newcommand{\nuc}{\nu_{\rm c}}
\newcommand{\ncc}{\nu_{\rm c,9}}

\newcommand{\Bc}{B_{ d,15}}
\newcommand{\lac}{L_{A,44}}
\newcommand{\Bfms}{B_{\rm FMS}}

\begin{document}
\begin{CJK*}{UTF8}{gkai}

\title{An Intermediate-Field Fast Radio Burst Model and the Quasi-Periodic Oscillation}
\volnopage{Vol.0 (20xx) No.0, 000--000}      
\setcounter{page}{1}          
   
\author{Jie-Shuang Wang \inst{1}
\and  Xinyu Li \inst{2,3}
\and  Zigao Dai \inst{4,5}
\and Xuefeng Wu \inst{6}
}

\institute{Max-Planck-Institut f\"ur Kernphysik, Saupfercheckweg 1, D-69117 Heidelberg, Germany, jswang@mpi-hd.mpg.de\\
\and
Canadian Institute for Theoretical Astrophysics, 60 St George St, Toronto, ON M5R 2M8
\and
Perimeter Institute for Theoretical Physics, 31 Caroline Street North, Waterloo, Ontario, Canada, N2L 2Y5
\and
Department of Astronomy, University of Science and Technology of China, Hefei 230026, China
\and
School of Astronomy and Space Science, Nanjing University, Nanjing 210023, China
\and 
Purple Mountain Observatory, Chinese Academy of Sciences, Nanjing 210023, China
}

\abstract{\\
Quasi-periodic oscillation (QPO) signals are discovered in some fast radio bursts (FRBs) such as FRB 20191221A, as well as in the X-ray burst associated with the galactic FRB from SGR 1935+2154.
We revisit the intermediate-field FRB model where the radio waves are generated as fast-magnetosonic waves through magnetic reconnection near the light cylinder. 
The current sheet in the magnetar wind is compressed by a low frequency pulse emitted from the inner magnetosphere to trigger magnetic reconnection.
By incorporating the wave dynamics of the magnetosphere, we demonstrate how the FRB frequency, the single pulse width, and luminosity are determined by the period, magnetic field, QPO frequency and quake energetics of the magnetar. 
We find that this model can naturally and self-consistently interpret the X-ray/radio event from SGR 1935+2154 and the QPO in FRB 20191221A. 
It can also explain the observed wide energy range of repeating FRBs in a narrow bandwidth. 
}

\titlerunning{An Intermediate-Field FRB model and QPO}
\authorrunning{Wang et al.}

\maketitle

\section{Introduction}

In recent years, the study of fast radio bursts (FRBs) has been greatly advanced by the progress in the observations. 
New detections of the galactic events \citep{CHIME2020Naturmagnetar,Bochenek2020Natur}, burst polarisation \citep[e.g.][]{Michilli2018,Luo2020Natur}, burst morphology \citep[e.g.][]{Pleunis2021ApJ}, source periodicity \citep[e.g.][]{Chime2020NaturPeriod,Pastor-Marazuela2021Natur,CHIME2022Natur}, source activity \citep[e.g.][]{Li2021Natur} and host galaxies \citep[e.g.][]{Bassa2017ApJL,Bhandari2022AJ}, have led to constraints on both progenitor models and radiation mechanisms.
However, a complete theoretical understanding of fast radio bursts is still not available and requires more effort \citep[see][for recent reviews]{Zhang2020Natur,Xiao2021SCPMA}. 

Highly magnetised compact objects are usually involved as the central engine of FRBs, such as magnetars, pulsars, and accreting black holes. Among them, the magnetar scenario has been confirmed by the detection of the galactic event \citep[e.g.][]{CHIME2020Naturmagnetar,Bochenek2020Natur,Li2021NatAs,Mereghetti2020ApJL,Ridnaia2021NatAs,Tavani2021NatAs}. 
Theoretically, the energy resource can come from the internal magnetic energy of magnetars \citep[e.g.][]{Popov2013,Katz2016,Beloborodov2017,Margalit2020ApJL} or the gravitational potential energy when a companion star is involved \citep[e.g.][]{Geng2015,Wang2016,Dai2016,Wang2018,Zhang2020ApJL,Dai2020ApJL,Most2022MNRAS}. 

Currently, the mainstream FRB radiation mechanisms can be divided into two categories: the near-field (or close-in) and the far-field (or far-away) model. 
The near-field model, mainly based on the coherent curvature radiation model near the magnetar surface \citep[e.g.][]{Kumar2017,Yang2018,Lu2020MNRAS,WangFY2020ApJ,Yang2020ApJL,WangWY2020ApJ}, can explain the complex temporal behavior of FRBs, while the bunching mechanism for the coherent curvature radiation is unclear. 
Recently, it is found that magnetospheric radio waves suffer strong dissipation when the wave becomes non-linear and can not escape from the magnetosphere \citep{Beloborodov2022PRL,Chen2022arXiv}. 
The far-field model incorporates the synchrotron maser emission at the shock front far from the magnetosphere as the coherent radio emission mechanism  \citep[e.g.][]{Lyubarsky2014,Beloborodov2017,Margalit2018,Waxman2017,Metzger2019MNRAS,Beloborodov2020ApJ,Plotnikov2019,Yu2020,Margalit2020ApJL,Wu2020ApJL,Xiao2020ApJL}.
This mechanism has been demonstrated using kinetic plasma simulations \citep[e.g.][]{Plotnikov2019,Sironi2021PhRvL}.
However, it is difficult to explain the sub-second scale quasi-periodic oscillations (QPOs) from FRBs \citep{CHIME2022Natur,Pastor-Marazuela2022arXiv}. 
It is also found that neither model can fully explain the X-ray/radio event from SGR 1935+2154 \citep{Wang2020ApJ}. 

Recently, a new type of FRB model is proposed where an FRB is radiated as fast-magnetosonic (FMS) waves generated from violent magnetic reconnection triggered by a low-frequency pulse (LFP) compressing the current sheet \citep{Lyubarsky2020ApJ}. 
As the coherent radio emission is produced near the light cylinder, we refer to it as the ``intermediate-field'' model.
Kinetic plasma simulations \citep{Philippov2019,Mahlmann2022ApJL} have successfully demonstrated the emission of coherent FMS waves through this mechanism. 
It is also found that the high linear polarisation \citep{Lyubarsky2020ApJ} and the downward frequency drifting \citep{Mahlmann2022ApJL} can be well explained. 
However, the predicted frequency is significantly lower than the observation of the SGR 1935+2154 event \citep{Wang2020ApJ}.

In this paper, we revisit the intermediate-field model and propose that the LFP is produced through nonlinear conversion of Alfv\'en waves, and propose a toy model for the generation of QPOs. 
In section \ref{sec:radiation}, we briefly review the coherent radiation mechanism from magnetic reconnection.
In section \ref{sec:FRB_QPO}, we study the injection of energy through Alfv\'en waves, the wave dynamics in the magnetosphere, and the generation of FRBs and QPOs. 
In section \ref{sec:explainQPO}, we apply our model to observations. 
The conclusion and discussion are presented in section \ref{sec:conclusion}.
Throughout this paper we adopt the shorthand $X=X_n\times 10^n$ to describe the normalization of quantity $X$ in cgs units.

\section{Reconnection Driven Coherent Radio Emission}\label{sec:radiation}


Magnetic reconnection is a process of changing the magnetic topology when two oppositely directed magnetic field lines approach each other forming a current sheet at the centre.
The onset of magnetic reconnection is triggered when the current sheet becomes tearing or kink unstable and breaks into a self-similar chain of plasmoids extending down to the kinetic scale \citep{Uzdensky2010PhRvL}.
Fast magnetic reconnection proceeds when the plasmoids collide and merge into larger islands.
During the coalescence of plasmoids, magnetic energy is dissipated and FMS waves are produced.
In highly magnetized systems with magnetization parameter $\sigma\equiv B^2/4\pi \rho c^2 \gg1$, FMS waves have similar dispersion relation as vacuum electromagnetic waves \citep{Thompson1998,Li2019} and can convert to coherent radio waves \citep[e.g.][]{Lyubarsky2019,Philippov2019,Lyubarsky2020ApJ}.

It has been proposed that the coherent emission of FRBs are GHz FMS waves generated by the plasmoid collision during magnetic reconnection \citep{Lyubarsky2020ApJ}.
In their picture, the current sheets in the magnetar wind are compressed by an external outgoing LFP and become tearing unstable under the perturbation, and magnetic reconnection is  initiated.
Coherent FMS waves are produced in the reconnection and can escape as fast radio bursts.
Kinetic simulations have confirmed this process and found that FMS waves takes only a small fraction of the reconnection energy \citep{Mahlmann2022ApJL}.

The characteristic frequency of the coherent FMS waves is determined by the size of the plasmoid $\lambda'_c$. Note we use the primed symbols to mark parameters measured in the co-moving frame of the plasma throughout the paper.
In the plasma comoving frame, $\omega'_c \approx c/ \lambda'_c$.
The typical plasmoid size is found to be proportional to the width of the current sheet $\lambda'_c=\varsigma a'$ with $\varsigma\sim10-100$ \citep{Philippov2019,Lyubarsky2020ApJ}, and the width depends on the Larmor radius $a'=\varpi r_{\rm L}$ with $\varpi$ being a few. 

The plasmoid size can be obtained by balancing the pressure and energy \citep{Lyubarsky2020ApJ},
\begin{eqnarray}
    \lambda'_c=\sqrt{\frac{8\pi\beta_{\rm rec}\varpi}{e\sigma_T}}\frac{m_ec^2}{B'^{3/2}},
\end{eqnarray}
where $\sigma_T$ is the Thomson scattering cross section, $\beta_{\rm rec}\sim0.1$ is the reconnection rate, $B'=B/\Gamma$ is the magnetic field in the comoving frame and $\Gamma$ is the bulk Lorentz factor. 
Therefore, the characteristic frequency in the observer's frame is,
\beq
\nu_{\rm c}\equiv\Gamma\omega'_c/2\pi\approx 2.1 \eta_1^{-1} \Gamma^{-1/2} B_{8}^{3/2}~{\rm GHz},\label{Eq:nuc}
\eeq
where $\eta\equiv\varsigma \varpi^{1/2}\beta_{\rm rec}^{1/2}$. 
This expression is confirmed numerically but with a slightly different expression of $\eta$ \citep{Mahlmann2022ApJL}. 

\begin{figure}
    \centering
    \includegraphics[width=0.6\textwidth]{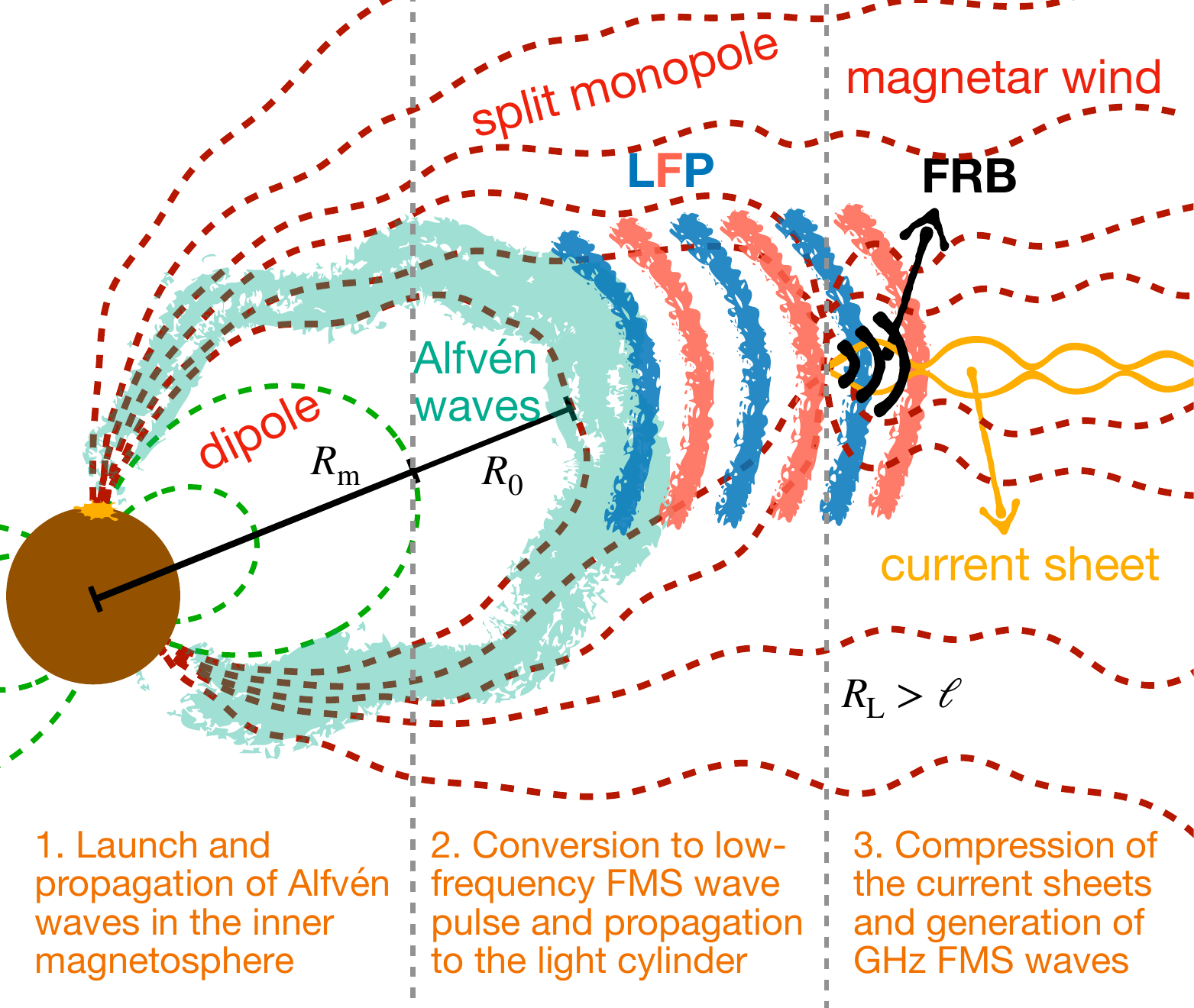}
    \caption{The schematic picture of the intermediate-field FRB model (not to scale): 
    After launch, Alfv\'en waves propagate along field lines of $R\gtrsim R_0$, and become non-linear at $R_{\rm m}$, which leads to the conversion to an LFP consisting of low-frequency FMS waves. 
    (The Alfv\'en waves and LFP are drawn with different styles and unrealistic wavelengths to make their presence clear.)
    The LFP propagates through the light cylinder and compresses the current sheet in the striped magnetar wind.
    GHz FMS waves are generated by merging islands during the violent magnetic reconnection, which can escape as an FRB.}
    \label{fig:magnetar_model}
\end{figure}

\section{Intermediate-field FRB models and QPOs}\label{sec:FRB_QPO}

Our model is illustrated in Fig. \ref{fig:magnetar_model}, which describes how an FRB is produced from a magnetar quake: (1) Alfv\'en waves are generated by magnetar quakes, which propagates in the inner magnetosphere. 
(2) When the wave amplitudes becomes comparable to the background magnetic field, Alfv\'en waves can convert to FMS waves with the same wave frequency. The low-frequency FMS waves compose an LFP and propagates into the striped wind. 
(3) The LFP compresses the current sheets in the striped wind and initiate violent magnetic reconnection, which can generate GHz FMS waves.
(4) Current sheets are also build up behind the LFP, which can generate high-frequency FMS waves but suffers strong dissipation.
In the following subsections, we discuss these four points in details respectively.

\subsection{Launch and propagation of Alfv\'en waves in the inner magnetosphere}

For a magnetar with a surface dipole magnetic field $B_d$ and radius $R_*\approx10^6$\,cm, the total magnetic energy is $E_{B}=B_d^2R_{*}^3/6\approx1.7\times10^{47}\Bc^2$\,erg. 
The internal magnetic field $B_*$ can be higher and we parameterize it by $B_*= \zeta B_d$ with $\zeta>1$.
The internal magnetic energy can be released by a sudden starquake and crustal motion which are accompanied by the crustal shear oscillations \citep[e.g.][]{Thompson1995MNRAS,Thompson2001,Duncan1998ApJL}.
Such oscillations are proposed to explain the QPOs observed in magnetar X-ray flares \citep[e.g.][]{Israel2005,Strohmayer2005,Strohmayer2006ApJ}.

The oscillation frequency depends on the magnetar properties. 
It has been found that the wave frequency of the toroidal shear modes with the lowest radial node (symbolized by $_lt_0$) follows $\fa(_lt_0)\approx12.2l^{1/2}(l+1)^{1/2}\,\rm{Hz}$ with $l\geq2$ being an integer in the magnetar crust \citep{Duncan1998ApJL}. 
Later it is found that the crustal shear modes will couple with the MHD modes of the magnetar core, and the frequency and duration of the oscillation will be modified \citep[e.g.][]{Levin2007MNRAS,Hoven2012,Gabler2016MNRAS}. 
Such theories predict that the fundamental frequency of magnetar QPOs can be as low as $\fa\sim\mathcal{O}(1)$ Hz, and it can be much higher for high order overtones, even up to $\fa\sim\mathcal{O}(10^3)$ Hz \citep[e.g.][]{Hoven2012}. 
The power spectrum of QPOs is found to be dominated by several frequencies depending on the magnetar properties, e.g. Fig. 9 in \cite{Hoven2012}, which is in general consistent with observations such as Fig. 3 in \cite{Israel2005}.

As the shear wave propagates in the crust, it shakes the anchored magnetic field lines, and launches Alfv\'en waves into the magnetosphere \citep{Thompson2001,Li2016,Thompson2017,Bransgrove2020}. 
The Alfv\'en waves carry energy into the magnetosphere and propagate along the field lines. 
The field-line equation is $r=R\sin^2{\theta}$, where $R$ is the maximum distance of the field line to the magnetar center and $\theta$ is the angle to the magnetic axis. 
For field lines with $R\gg R_*$, the field-line length is $l\simeq\pi R/2$.

The generated Alfv\'en waves of frequency $\fa\sim 1-10^3$ Hz in the magnetosphere can propagate along field lines with $l\geq l_0=\pi R_0/2=c/\fa=1.9\times10^{10}\fa^{-1}~{\rm cm}$ as a perturbation. 
Note that Alfv\'en waves travelling on field lines with $l<l_0$ may suffer strong nonlinear interaction and the dissipation mechanism is unclear. Thus we only consider magnetar quakes occurred near the magnetic polar region, so that magnetar oscillations can launch Alfv\'en waves into the magnetosphere before damping \citep[e.g.][]{Levin2007MNRAS,Hoven2012}.

We consider a sudden release of internal magnetic energy of $E_q=h S B_*^2/8\pi$ in the crust with a crust thickness $h\approx10^5$\,cm and an area $S_q$. 
Note in the quake region, the internal magnetic energy can be much larger than the surface field with $B_*/B_d=\zeta\gg1$.
The magnetar quake can simultaneously launch a number of waves with different frequencies, that are of different damping time-scales \citep[e.g.][]{Levin2007MNRAS,Hoven2012,Gabler2016MNRAS}. 
We consider that the Alfv\'en wave of frequency $f_A$ is the dominated one which carries most of the energy, and a number ($n_A$) of waves at this frequency has been excited. 
The energy carried by each wave is parameterized as $\epsilon E_q\approx L_A/f_A$ with a power $L_A$ and $\epsilon<1/n_A$. 
The wave is launched from the magnetar surface within the same area, 
\begin{eqnarray}
  S_q=2.5\times10^8\lac\fac^{-1}\Bc^{-2}\epsilon^{-1}\zeta^{-2}\,{\rm cm}^2.
\end{eqnarray}
The corresponding wave amplitude in the magnetosphere is ${\delta B}=\sqrt{8\pi L_A/S_qc}$.
Its relative amplitude to the background field near the magnetar surface is 
\begin{eqnarray}
    \frac{\delta B}{B_d}\approx 0.02\fac^{1/2}\epsilon^{1/2}\zeta,
\end{eqnarray} 

\subsection{Conversion to an LFP and its propagation in the outer magnetosphere}

As the Alfv\'en wave propagates in the magnetosphere, its amplitude grows as ${\delta B}/{B_d}\propto r^{3/2}$ \citep[e.g.][]{Kumar2020,Yuan2020}. 
When the relative amplitude grows to ${\delta B}/{ B}\approx1$ \citep{Yuan2020,Yuan2022ApJ} at a distance of
\begin{eqnarray}
   R_{\rm m}\approx 1.4\times 10^7\fac^{-1/3}\epsilon^{-1/3}\zeta^{-2/3}~\rm{cm},
\end{eqnarray}
the Alfv\'en waves will become nonlinear and shear the magnetosphere. 
The background magnetic field here can be estimated from a dipole $B_{\rm m}=B_*R_*^3R_{\rm m}^{-3}$.
Under the strong shear, the magnetosphere becomes unstable.
A large part of the Alfv\'en waves will be converted to FMS waves that propagates outwards radially in the form of a LFP \citep{Li2019,Li2021,Yuan2021} due to the interaction with the curved magnetic field line. 
The remaining Alfv\'en waves will dissipate inside the magnetosphere by nonlinear instabilities \citep{Yuan2020} or absorbed by the magnetar \citep{Li2015}.

We here mainly consider the LFP, that can carry a large portion of the energy of Alfv\'en waves \citep{Yuan2022ApJ}. 
The LFP is made up of low-frequency FMS waves converted from the propagating Alfv\'en waves.
This process can be viewed as the nonlinear interaction between Alfv\'en waves and the curved background magnetic field.
The leading order wave interaction can be treated as three-wave interaction with 
\begin{eqnarray}
    \boldsymbol{k}_A+\boldsymbol{k}_{\rm bg}=\boldsymbol{k}_{\rm LFP},~{\rm and} ~f_{\rm LFP}=f_A,\label{eq.fa=ffms}
\end{eqnarray}
where $\boldsymbol{k}$ is the wavenumber and $\boldsymbol{k}_{\rm bg}$ is the wavenumber of the background magnetic field \citep{Yuan2021}. 
Notably in this process the generated low-frequency FMS wave will have the same frequency as the Alfv\'en wave, as the background field does not have a time component, i.e. $f_{\rm bg}=0$. 
The LFP will propagate outward radially with its thickness conserved as found by numerical simulations \citep{Yuan2022ApJ}. Therefore the LFP keeps its wave frequency when propagating.

During the propagation of the LFP at $r>R_{\rm m}$, its toroidal field follows $rB_{\rm LFP } = B_{\rm LFP, m}R_{\rm m}$ with $B_{\rm LFP, m}\approx B_{\rm m}$ \citep{Parfrey2013,Yuan2020,Yuan2022ApJ}.
As the energy contained in the LFP is higher than that of background field, the magnetosphere is distorted. 
The poloidal field follows $r^2 B_p= B_{\rm m}R_{\rm m}^{2}$ from the conservation of magnetic flux. 
Thus the magnetosphere at $R>R_{\rm m}$ can be treated as a split monopole.
When the LFP reaches the light cylinder at $R_L=c P/2\pi$, where $P$ is the magnetar rotation period, its toroidal field and the poloidal field of the magnetosphere are given by
\begin{eqnarray}
    B_{\rm LFP, L}=3.4\times10^8\Bc\fac^{2/3}\epsilon^{2/3}\zeta^{4/3}\left(\frac{P}{3\,{\rm s}}\right)^{-1}\,\rm{G}\label{eq:bphi_LC};\\
    B_{p,m}(R_{\rm L})=3.4\times10^5\Bc\fac^{1/3}\epsilon^{1/3}\zeta^{2/3}\left(\frac{P}{3\,{\rm s}}\right)^{-2}\,\rm{G}.\label{eq:bpol_LC}
\end{eqnarray}

When propagating in the magnetosphere, the FMS waves also suffer from nonlinear steepening when the wave amplitude is comparable to the background magnetic field, and form shocks \citep{Lyubarsky2020ApJ,Beloborodov2022PRL,Chen2022arXiv,Beloborodov2022arXiv}. 
Considering the nonlinear steepening caused by the variation of wave velocity across the wavelength, which depends on the local density and magnetic field \citep{Lyubarsky2020ApJ}, the shock formation distance is 
\begin{eqnarray}
    R_{\rm stp}\sim \sigma_{\rm LFP} \Gamma^2 c/f_A= 3\times10^{10}\Gamma_1^2\sigma_{\rm LFP}~{\rm cm},
\end{eqnarray}
where the plasma bulk Lorentz factor for the LFP in the magnetosphere could be mild relativistic $\Gamma\sim10$.
Thus as long as the magnetization of the LFP is $\sigma_{\rm LFP}\gg1$, the shock formation distance is outside the light cylinder for magnetars.
However, the effect of high electric field may still be able to dissipate a fraction of LFP energy \citep{Beloborodov2022arXiv}, which is not considered here.

\subsection{Generation of FRBs and QPOs in the compressed current sheets in the striped wind}

The LFP will eventually enter the striped wind and accelerates the plasma to a bulk Lorentz factor of $\Gamma\approx0.5\sqrt{B_{\rm LFP, L}/B_{\rm w}(R_{\rm L}) }$ by compressing it, where the wind field is described as a monopole with $B_{\rm w}(r)=B_{p}(R_L)R_L/r$ and $B_{p}(R_L)$ is the poloidal field of the magnetosphere at the light cylinder. 
The pulse front propagates into the magnetar wind launched by the unperturbed magnetosphere with $B_{p}(R_L)=B_*R_*^3R_{\rm L}^{-3}$.
However, the energy carried by the LFP amplifies the magnetic field at the light cylinder to $B_{p,m}(R_{\rm L})$, which further enhances the toroidal magnetic field in the striped wind to $B_{\rm w}(R_{\rm L})\sim B_{p,m}(R_{\rm L})$. 
Therefore, the rear part of LFP will interact with the current sheets in the striped wind launched by the perturbed magnetosphere (see. Fig. \ref{fig:magnetar_model}). 
In reality, the amplification of the wind toroidal field and the formation of the current sheet are highly dynamic, we therefore expect the wind field to be in the range $B_{\rm w}(R_{\rm L})\in [B_*R_*^3R_{\rm L}^{-3},\; B_{p,m}(R_{\rm L})]$.

The corresponding maximum and minimum Lorentz factors of the accelerated LFPs are then
\begin{eqnarray}
    &&\Gamma^{\rm max}=496 \epsilon^{1/3}\zeta^{2/3}\fac^{1/3}\left(\frac{P}{3\,{\rm s}}\right);\nonumber\\
    &&\Gamma^{\rm min}=16 \epsilon^{1/6}\zeta^{1/3}\fac^{1/6}\left(\frac{P}{3\,{\rm s}}\right)^{1/2}.\label{eq.Gamma}
\end{eqnarray}
In the following we parameterize $\Gamma=100\Gc$ which lies between $\Gamma^{\rm min}$ and $\Gamma^{\rm max}$. 
And we expect $\Gamma\rightarrow\Gamma^{\rm min}$ for LFPs consisting of multiple waves (e.g. $n_A\gtrsim3$), such as in the SGR 1935+2154 event and FRB 20191221A (see Sec. \ref{sec:explainQPO} for more details).

Magnetic reconnection will be triggered when the LFP compresses the current sheet in the striped wind, which leads to the generation of high-frequency FMS waves as discussed in Section \ref{sec:radiation}.
Substituting $B_{\rm LFP,L}$ and $\Gamma$ into Eq.~(\ref{Eq:nuc}), we obtain the frequency of the FMS waves,
\begin{eqnarray}
    \nuc=1.3\fac\Bc^{3/2}\epsilon\zeta^2\left(\frac{P}{3\,{\rm s}}\right)^{-3/2}\eta_1^{-1}\Gc^{-1/2}\,\rm{GHz},\label{eq:nu_final}
\end{eqnarray}
which lies in the radio band.

The overall energy dissipation in the magnetic reconnection will be mediated by the wave frequency of LFP, which is identical to the Alfv\'en wave frequency (Eq. \ref{eq.fa=ffms}). In this case, we expect 
\begin{eqnarray}
    \ffrb\sim f_{\rm LFP}=\fa,
\end{eqnarray}
which makes the observed QPO signals in some FRBs.
Note in this expression, we assume $\fa>1/P$, which will in general be satisfied for magnetars with $P\gtrsim0.1$\,s. Otherwise the FRB profile will also be modulated by the spin period. 

As the plasma is accelerated to a bulk Lorentz factor of $\Gamma$, the observed width of one single pulse and peak luminosity will be 
\begin{eqnarray}
    W_{\rm p}\approx1/(\Gamma\fa),\label{eq:tpulse}\\
     L_{\rm FMS}\approx\kappa \Gamma^2L_A \label{eq:lfrb}.
\end{eqnarray}

The produced FMS waves suffers from nonlinear wave interactions, and it can escape only when its optical depth $\tau_{\rm NL}\lesssim10$ \citep{Lyubarsky2020ApJ}, which is given by
\begin{eqnarray}
    \tau_{\rm NL}\sim \frac{\kappa 2\pi\nu R_{\rm L}}{2\Gamma^2c}=0.5\kappa_{-5.5}\ncc\left(\frac{P}{3\,{\rm s}}\right)\Gc^{-2},
\end{eqnarray}
where $\kappa=\Bfms^2/B_{\rm LFP, L}^2$ is the energy ratio of the high-frequency FMS waves to the LFP. In 2D simulations, it is found that a fraction $\kappa\sim10^{-4}$ of energy from the LFP is converted to the FMS, interpreted as an upper limit of the efficiency \citep{Mahlmann2022ApJL}. 
The ratio of radio to X-ray energy of FRB200428 is found to be $10^{-6}-10^{-5}$ \citep{Ridnaia2021NatAs}. 
Here we take $\kappa\sim10^{-5.5}$ to be the fiducial value. 
Therefore, the FMS waves generated at the light cylinder can escape without significant dissipation and be observed as FRBs.

\subsection{The current sheets behind the LFP}

There are also current sheets behind the LFP inside the magnetosphere, which has been suggested to produce an FRB \citep{Wang2020ApJ,Yuan2020,Yuan2022ApJ}. 
The reconnected fields are the poloidal component \citep{Yuan2022ApJ}, and the frequency of FMS waves produced in those current sheets is 
\begin{eqnarray}
    \nu=4.0\times 10^{14} \Bc^{3/2} \fa^{1/2}\zeta\Gamma_{\rm pl} \epsilon^{1/2} (R_{\rm m}/R)^{3} \eta_1^{-1} ~{\rm Hz},
\end{eqnarray}
where the plasmoid bulk Lorentz factor inside the magnetosphere can be a few, $\Gamma_{\rm pl}\sim1-10$.
The FMS wave frequency ranges from optical to radio for $R\in(R_{\rm m}, R_{\rm L})$ for the current sheet inside the magnetosphere. 

However, the FMS waves emitted deep inside the magnetosphere at $R\sim R_{\rm m}$ also suffers from strong dissipation of the nonlinear wave interactions.
The optical depth inside the magnetosphere is $\tau_{\rm NL}\approx({\Bfms}/{B_p})^2\omega R/c$ \citep[see Eq. A3 in][]{Lyubarsky2020ApJ} with the FMS wave magnetic field ${\Bfms}=\sqrt{2L_{\rm FMS}/ r^2}$ and $B_p$ being treated as a monopole.
Setting $\tau_{\rm NL}\sim10$, we obtain a mean-free-path 
\begin{eqnarray}
\ell\sim3.4\times10^6\Bc^{1/2}\fac^{1/6}\epsilon^{1/6}\zeta^{1/3}\nu_{14}^{-1/4}L_{\rm FMS,42}^{-1/4}\,{\rm cm},
\end{eqnarray} 
which is much smaller than $R_L$ for a typical magnetar with period $0.1-10$~s.
Therefore, such optical radiation will be dissipated inside the magnetosphere. 

%

The frequency of the FMS waves produced by the current sheet behind the LFP at larger radii ($R\sim R_L$) could lie in the radio band.
However, as the magnetic reconnection is supported by the reversed poloidal magnetic field, the magnetic flux is much smaller at near the light cylinder,
\begin{eqnarray}
    \frac{B_{p,m}^2(R_{\rm L})}{B_{\rm LFP, L}^2}=1.0\times10^{-6}\fac^{-2/3}\epsilon^{-2/3}\zeta^{-4/3}\left(\frac{P}{3\,{\rm s}}\right)^{-2}.
\end{eqnarray}
Therefore, the FMS waves produced behind the LFP cannot be the primary source of FRB, as its available magnetic energy here is much smaller than the energy of the LFP.

\section{Application to the observed QPOs of FRBs and the SGR 1935+2154 event}\label{sec:explainQPO}

We now apply our model to the observed QPOs in the X-ray/radio burst of SGR 1935+2154 and in other FRBs.
Observations show that the FRB from the galactic magnetar, SGR 1935+2154 \citep{CHIME2020Naturmagnetar,Bochenek2020Natur}, is associated with a hard X-ray burst by several X-ray instruments\citep{Li2021NatAs,Mereghetti2020ApJL,Ridnaia2021NatAs,Tavani2021NatAs}. 
This magnetar is of spin period and surface magnetic field $P=3.2$\,s and $B_*=2.2\times10^{14}$\,G respectively \citep{Olausen2014,Israel2016}. 
Two radio pulses separated by $30$\,ms are found in this FRB, while more pulses are detected in X-ray bands. 
The reason may be that the opening angle of the radio emission is much smaller that that of the X-ray emission \citep[e.g.][]{Wang2020ApJ}.
Especially, a QPO of $\fa\approx40$\,Hz and $n_A\gtrsim8$ is detected in the X-ray light curve by the $Insight$-HXMT \citep{Li2022ApJ}. 
The observed pulse width is $0.3-0.6$\,ms by CHIME and STARE2 in $0.6-1.3$\,GHz \citep{CHIME2020Naturmagnetar,Bochenek2020Natur}.
Substituting $\epsilon<1/n_A\sim1/8$, $W_{\rm p}\sim0.5$\,ms, and  $\nuc\sim1$\,GHz as well as the observed radio peak luminosity $L_{\rm FMS}\approx10^{38}$\,erg/s \citep{Bochenek2020Natur,Zhou2020ApJ} into Eqs. (\ref{eq:nu_final}, \ref{eq:tpulse} and \ref{eq:lfrb}), we obtain $\Gamma\sim50$, $\zeta\gtrsim3.5\eta_1^{1/2}$ and $L_A\approx 1.4\times10^{40} \kappa_{-5.5}^{-1}$\,erg/s.
The required $\Gamma$ is in the allowed range $[22\eta_1^{1/6}, 977\eta_1^{1/3}]$ from Eq. (\ref{eq.Gamma}), which is close to $\Gamma^{\rm min}$ as we expected. 
And the required $L_A$ is comparable to the observed X-ray luminosity, $10^{40}$~erg/s \citep{Mereghetti2020ApJL,Li2021NatAs}. 
Therefore, our model can explain this event self-consistently.

QPO signals are also observed in individual FRB events, such as the $4.6$\,Hz QPO in FRB 20191221A, $357.1$\,Hz in FRB 20210206A, $93.4$\,Hz in FRB 20210213A detected by CHIME \citep{CHIME2022Natur}, and the $2.4$\,kHz QPO in FRB 20201020A detected by Apertif \citep{Pastor-Marazuela2022arXiv}. 
It has been suggested that the $4.6$\,Hz QPO may be caused by the magnetar spin period  \citep{Beniamini2022arXiv221107669B}, however, such a spin-period scenario will be difficult to explain the high-frequency QPOs, especially the $2.4$\,kHz one \citep{Pastor-Marazuela2022arXiv}. While the magnetar oscillation can offer a natural explanation.
The observed QPOs with frequencies larger than $93.4$\,Hz mentioned above can be explained by the overtones in a wide parameter range. While the $4.6$\,Hz QPO in FRB 20191221A may relate to the fundamental mode. We here focus on this source, which has the highest significance \citep{CHIME2022Natur}. 
The observed radio frequency, single pulse width, and QPO frequency of FRB 20191221A are $\nuc\approx0.7$\,GHz, $W_{\rm p}\approx4$\,ms, and $f_A\approx4.6$\,Hz with $n_A\approx9$, respectively. 
The total duration of this event is $\sim3$\,s and we would expect $P>3$\,s so that the QPO signal is not contaminated by the magnetar rotation. 
Substituting them into Eqs. (\ref{eq:nu_final} and \ref{eq:tpulse}), we obtain $\Gamma\approx54$ and $\zeta\gtrsim9\eta_1^{1/2}(P/3\,{\rm s})^{3/4}\Bc^{-3/4}$. 
The required $\Gamma$ is in the allowed range $[14\eta_1^{1/6}(P/3\,{\rm s})^{3/4}\Bc^{-1/4},367\eta_1^{1/3}(P/3\,{\rm s})^{3/2}\Bc^{-1/2}]$ and we find again that the derived bulk Lorentz factor is close to $\Gamma^{\rm min}$.
Thus the QPO in FRB 20191221A can also be explained by our model.

\section{Conclusion and Discussion}\label{sec:conclusion}

In this paper, 
we generalized the wave dynamics in the magnetosphere from the previous simulations results in \cite{Yuan2020,Yuan2022ApJ} and applied it in the intermediate-field FRB model. 
We focused on the recently observed QPOs in some FRBs and found that it can be self-consistently explained in the revised model.
The launch of Alfv\'en waves into the magnetosphere generates an LFP, which is made up of FMS waves at approximately the same frequencies of Alfv\'en waves ($\fa$). 
This further generates FRBs when the LFP dissipates its energy at the current sheet near the light cylinder. 
The FRB light curve is thus modulated by the frequency of the LFP, exhibiting QPOs at a frequency $\sim\fa$. 
The major difference of our calculations from \cite{Lyubarsky2020ApJ} is the treatment of magnetic field configurations of the LFP and the magnetar wind, as we considered the wave dynamics in the magnetosphere.

The FRB frequency (Eq. \ref{eq:nu_final}), the single pulse width (Eq. \ref{eq:tpulse}), and luminosity (Eq. \ref{eq:lfrb}) depend mainly on the magnetar's period and magnetic field and the quake energetics and QPO frequency.
With physically reasonable values for the parameters, we find that this model can naturally and self-consistently interpret the observed frequency, pulse width, luminosity and QPO signal of the radio/X-ray event from SGR 1935+2154 and FRB 20191221A.

Our model can also naturally explain the observed wide energy span detected in a relative narrow frequency band, such as the broad energy range ($4\times10^{36}-8\times10^{39}$\,erg) observed in FRB 121102 at 1.25 GHz \citep{Li2021Natur}, as the frequency does not depend on the flare energy apparently. 
And in certain parameter space for young magnetars, the frequency is in the optical band, indicating the possibility of producing fast optical transients.

In general, QPOs will modulate the FRB light curve. Direct identification of QPOs may require at least around ten individual pulses ($n_A\gtrsim10$) in one event \citep[e.g.][]{CHIME2022Natur}. 
For the high-frequency QPOs ($\fa\gtrsim1$ kHz), the corresponding FRB pulses may overlap with each other, making it difficult to identify from the data. 
While for the low-frequency QPOs ($\fa\sim10$ Hz), one may only expect to detect them from magnetars with $P\gtrsim10/\fa\sim1$\,s so that the QPO signal is not contaminated by the magnetar spin. 
Beside, the small opening angles of FRBs caused by the relativistic beaming effect also make it challenging to detect multiple pulses in a single event. 
Thus, direct detection of QPOs will only be possible for some rare events. 
However,
for repeating FRBs, QPOs would affect the waiting time distribution even if only several pulses are presented in the observed individual events. 
The peak at $3.4$\,ms in the waiting time distribution of FRB 121102 may be such a case \citep{Li2021Natur}.
Another possibility is to detect QPOs from the counterparts of FRBs, such as the non-thermal X-ray burst in SGR 1935+2154, although the radiation mechanism in X-ray QPOs requires further detailed studies.

As only a small portion of energy is dissipated to power FRBs, the majority of the energy will be dissipated into X-rays or carried by the ejecta. 
The ejecta would further power an afterglow or a nebula \citep[e.g.][]{Waxman2017,Beloborodov2017,Margalit2018,WangLai2020}. 
As magnetar flares may eject a large fraction of mass \citep[e.g.][]{Granot2006ApJ}, high-energy cosmic rays can also be produced in such afterglows/nebulae.
Besides, such a magnetic reconnection process could also take place in neutron star mergers \citep[e.g.][]{Wang2018,Most2022MNRAS} and accreting black holes \citep[e.g.][]{Beloborodov2017ApJ,Sridhar2021MNRAS}, thus we might also expect such reconnection driven transients from neutron star mergers, X-ray binaries, or active galactic nuclei.

\acknowledgements
We thank the referee for helpful comments.
JSW thanks J. Kirk, B. Reville, and F. Guo for discussions.
JSW acknowledges the support from the Alexander von Humboldt Foundation.
XL is supported by NSERC, funding reference \#CITA 490888-16 and the Jeffrey L. Bishop Fellowship. Research at Perimeter Institute is supported in part by the Government of Canada through the Department of Innovation, Science and Economic Development Canada and the Province of Ontario through the Ministry of Colleges and Universities.
ZGD is supported by the National Key Research and Development Program of China (grant No. 2017YFA0402600), the National SKA Program of China (grant No. 2020SKA0120300), and the National Natural Science Foundation of China (grant No. 11833003).
XFW is supported by the National Natural Science Foundation of China (Grant Nos. 11725314, 12041306) and the National SKA Program of China (2022SKA0130101).

\bibliographystyle{raa}
\bibliography{ref}
\end{CJK*}
\end{document}